# Parallel Implementation of Distributed Global Optimization (DGO)

Homayoun Valafar
Complex Carbohydrate Research Center
University of Georgia
220 Riverbend Road
Athens, GA 30602
homayoun@mond1.ccrc.uga.edu

Okan K. Ersoy
School of Electrical Engineering
Purdue University
West Lafayette, IN 47907
ersoy@ecn.purdue.edu

Faramarz Valafar
Complex Carbohydrate Research Center
University of Georgia
220 Riverbend Road
Athens, GA 30602
faramarz@mond1.ccrc.uga.edu

## Abstract

*Parallel implementations of distributed global optimization (DGO) [13] on MP-1 and NCUBE parallel computers revealed an approximate O(n) increase in the performance of this algorithm. Therefore, the implementation of the DGO on parallel processors can remedy the only draw back of this algorithm which is the $O(n^2)$ of execution time as the number of the dimensions increase. The speed up factor of the parallel implementations of DGO is measured with respect to the sequential execution time of the identical problem on SPARC IV computer. The best speed up was achieved by the SIMD implementation of the algorithm on the MP-1 with the total speedup of 126 for an optimization problem with n = 9. This optimization problem was distributed across 128 PEs of Mas-Par.*

## Introduction

The recent emergence of population based optimization techniques such as DGO or Genetic Algorithm (GA) have provided new means for the discovery of the global optimal points of a complex function. However, these algorithms greatly increase the execution time by producing a population of points which need to be evaluated by a complex function. This increase in the execution time has always been accounted as one of the disadvantages of these algorithms. In general, these algorithms exhibit an inherent parallelizable portion which can easily be implemented on parallel processors. The recent increase in the availability of parallel computers can be utilized to counteract the disadvantages of DGO or genetic algorithm.

Distributed global optimization (DGO) most resembles the Genetic Algorithm (GA)[1]. Like GA, Distributed Global Optimization (DGO) proceeds with the search for the optimal point by creating a population of points. Therefore DGO possesses most advantages of the GA. However unlike GA, DGO will not allow the unfit points to propagate to the next generation and thus the progression towards the optimal point is enhanced. Furthermore, the production of a child point is dependent on only the parent point and none of the other children. This will eliminate any inter-processor communication. The reduction in the communication overhead will result a great portion of the execution time to be occupied with computation. Computation is the exact section of these algorithms which can be distributed over n processors for a speed up of near O(n).

## Outline of DGO

DGO starts the search for the optimal point from a parent point. This parent point, is represented by n bits. 2n-1 new points will be generated by transformation of the parent point in a deterministic fashion.

DGO constructs and examines points in a tree like structure. A graphical illustration of the creation of population through an entire cycle is shown in Figure 1 (in this example a vector length of 4 bits is assumed). In the above graph, $S_i^j$ indicates the i-th element of the j-th stage. Each $S_i^j$ is a possible candidate for the next state S(i+1).

To study the search of the algorithm in the feasible region of space, we need to establish a notation for the search. The initial state of the algorithm is denoted as S(0). S(i) is referred to as the parent for the ith iteration. Furthermore, $S_i^j$ is the i-th element in the j-th stage

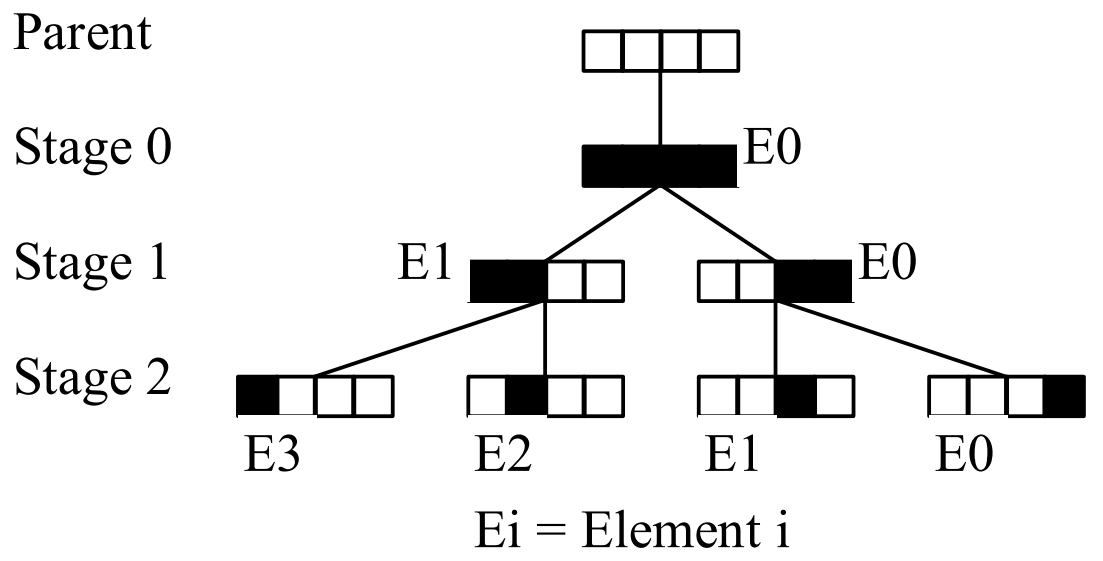

**Figure 1. An example of generation of population for n=4.**

The general outline of the algorithm is as follows:

1. Select an initial parent string (either the user's choice or a random point) and evaluate the function at that point.
2. From the current parent generate 2n-1 children by transforming different segments of the parent string.
3. Among these points find the one with the lowest function value.
4. If there is a new minimum, then set the new string as S(i+1) (the new parent point) and go to 2.
5. If there is no deeper minimum, then increase the resolution.
6. If the current resolution is less than the maximum resolution, then go to 2, else terminate the algorithm.

Although in step 2 any qualified transformation can be used, in this experiment a combination of gray code and binary inversion was used as the transformation . An effective transformation constructs the heart of this algorithm. For example, although binary inversion alone can be used as the transformation, the results of the optimization will not be effective. For more details on the gray code and inverse gray code transformations refer to [13]. In step 2 the transformation is constructed of the following steps:

1. Convert the entire binary string from two's complement to gray code.
2. Apply a binary inversion to the corresponding segment of the string.
3. Convert the string back to two's complement by the means of inverse-gray code transformation.

The above algorithm can be used for multidimensional problems as well by representing the set of all the independent variables as a single binary vector. Since all members of the population are created from one parent point, thus there are no dependencies in creation of each member. Conceptually, all members of a population can be created simultaneously. Because of this attribute of the algorithm, it can very easily be implemented on an SIMD machine for a total speedup of O(n).

## Theoretical and Practical Test Functions

Test functions for the one and two dimensional spaces are given in [1,2,7]. These test functions represent practical problems well. For comparison purposes the results of the new algorithm were compared to that of fmin (a function of matlab), Eureka , gradient descent[2,11], genetic algorithm and simulated annealing[3,4]. DGO was found to be the only algorithm which successfully discovered the global optimum point of each test function. Figures 2 and 3 display some examples of functions tested.

Although formulated test functions give some insight to the operation of a certain algorithm, the true test of success remains in the real world applications. Artificial Neural Network (ANN) [8,9,10,11] is one example of real life application which demands an effective optimization application. As a real test of DGO's capacity, it was applied to an ANN solving XOR [10,13] problem as well as 8 class remote sensing [10] application. The XOR problem contained 8 variables where as the remote sensing problem contained 688 variables. DGO proved itself a far better method of optimization for the learning of the neural network than the conventional gradient descent. These results are shown in Figures 4 and 5.

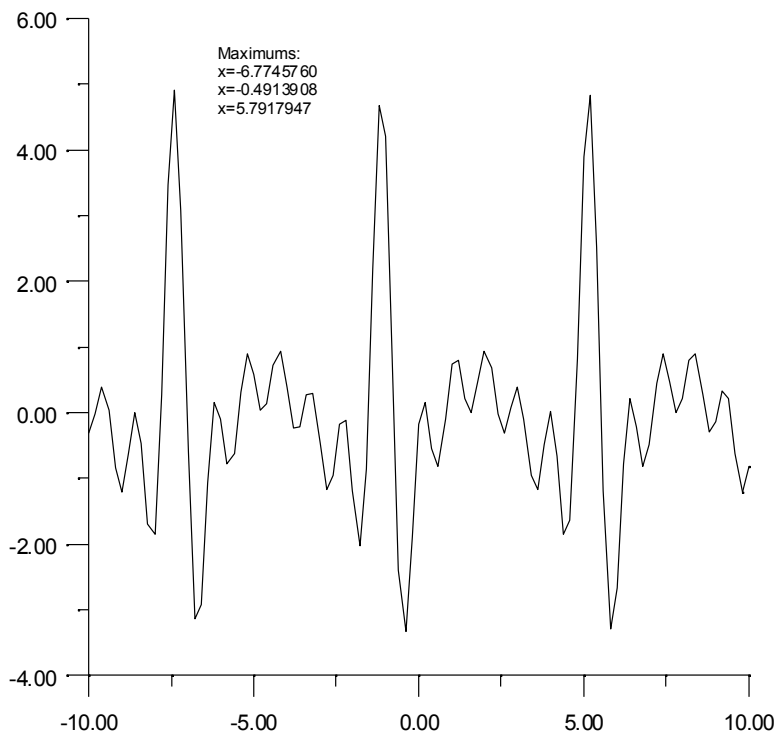

**Figure 2. A sample 2D test function.**

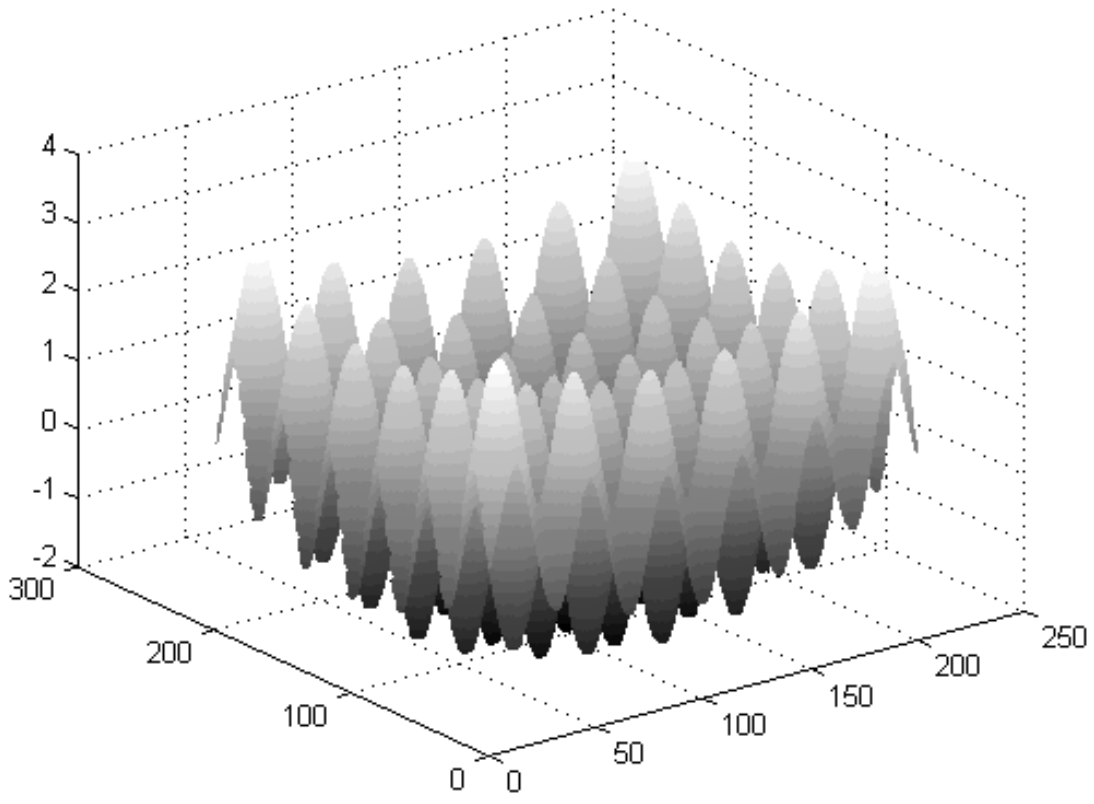

**Figure 3. A sample 3D test function.**

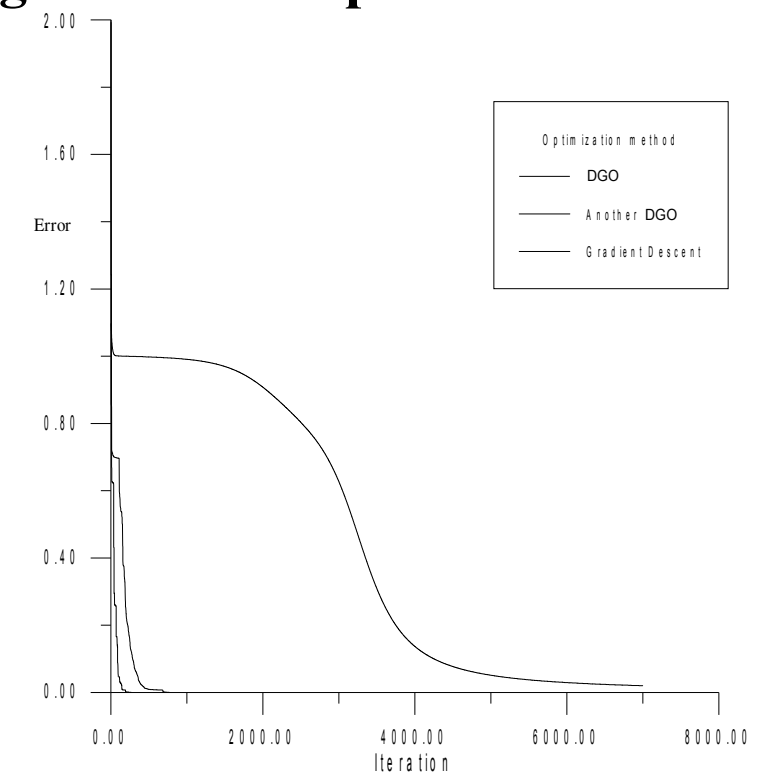

**Figure 4. Error trace of DGO and Gradient Descent.**

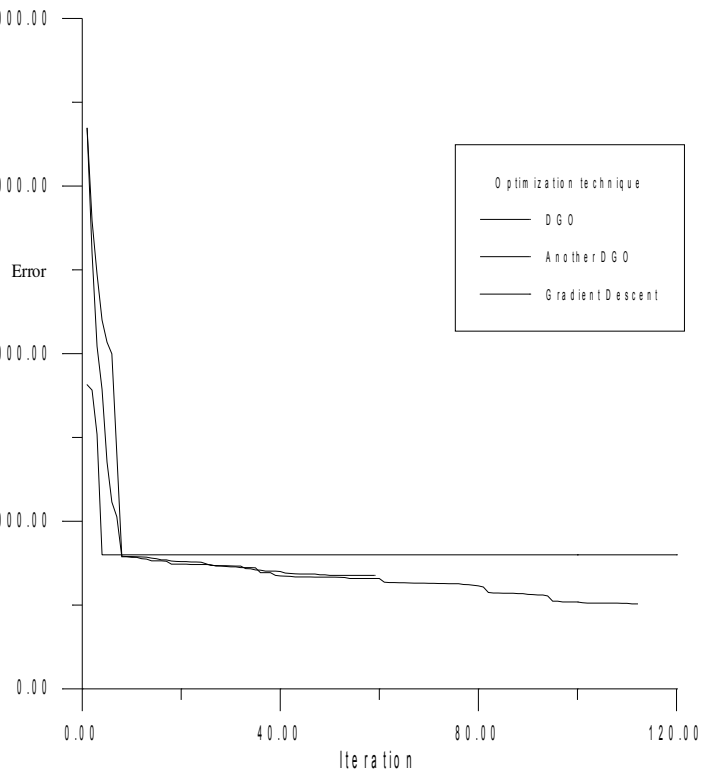

**Figure 5. Error trace of DGO versus Gradient Descent.**

## Parallel Implementation on MasPar (MP-1)

There are 16K number of PEs available on MP-1 which can be allocated to form different clusters. Each one of these clusters will start the optimization problem from their own starting point. Note that this clustering is efficient only if the problem size is small enough with respect to total number of PEs (16387 total number of PEs). The outline of the parallel code implementation is as follow:

1. ACU creates and distributes the initial state S(0) to 2n-1 PEs.
2. Each PE calculates its corresponding new point $s_i^j$.
3. Once the entire tree of $s_i^j$'s is constructed (with each element on one PE), a plural function evaluation results in the simultaneous evaluation of all the points.
4. The minimum of the values can be found using one of MP-1's library calls. In the current program the rank function was used.
6. The value of $s_i^j$ which corresponds to the minimum value can either be broadcasted to all other PEs or can be sent back to ACU. In this program the latter was selected. After ACU receiving the new $s_i^j$, it will proceed by distributing the new string to all PEs and then go to step 2
7. Continue until no better minimum is found.

Number of active PEs for each cluster is 2n-1 (where n is the vector size). If this number is small enough, then a number of clusters can be created for better performance.

## Parallel Implementation on Ncube

Although NCUBE is capable of having up to 1024 processors, the existing Ncube at Purdue contains 64 processors. This relatively small number of processors causes the parallel implementation to use virtual processing techniques [25, 3, 4]. The general outline of the algorithm is shown below:

1. PE0 creates and distributes the initial state S(0) to 2n-1 PEs (note that if n is a power of 2 then 2n will also be a power of two and fit within a hypercube).
2. Each PE calculates its corresponding new point $s_i^j$.
3. Once the entire tree of $s_i^j$'s is constructed among all PEs, a synchronized, parallel function evaluation results in the simultaneous evaluation of all these points.
4. The minimum of the values can be found by one of the following two methods:

   a. All PEs report their result to PE0. Then PE0 determines the smallest value.

   b. A method of cube reduction can be applied so that PE0 will end up with the minimum value.

   The fist method was implemented in our algorithm since usually the number of available PE's did not exceed 16. The second method is more effective for larger size cubes.

6. The value of $s_i^j$ which corresponds to the minimum value will be broad-caste to all other PEs. After each PE receives the new $s_i^j$, it proceeds with step 2
7. Continue until no better minimum is found.

As it was mentioned above, at each state there will be 2n-1 points generated and tested. If the number 2n-1 exceeds the number of PEs, then each PE must simulate the action of approximately $\left\lceil \frac{2n-1}{64} \right\rceil$ number of processors.

## Evaluation of Parallel Algorithms

The first step for the evaluation of the parallel algorithm is to establish the time complexity of the sequential code. The bulk of DGO algorithm consists of two nested for loops. The limits of each of these loops is static and a linear function of n (n is the number of variables). Figure 6 demonstrates the $O(n^2)$ complexity of DGO. An n dimensional quadratic function was selected as a generic bench marking function.

Four elements contribute toward the efficiency of a parallel code design. Each one of these components are listed and briefly described below.

- **Problem decomposition** means distributing a single problem across a number of processors.
- **Load balancing** refers to providing for approximately equal shares or computational load between the processors.
- **Communication** is required to allow a processor to receive information from another processor in order to perform its computation.
- **Synchronization** is keeping the computations synchronized across the multiple processors so that PEs do not waste time waiting for intermediate results from other PEs.

Based on the above four criteria it is easy to see how DGO would be an ideal algorithm for distribution across several PEs. DGO naturally decomposes in equal sections, requires very little inter PE communication and requires no semaphores for synchronization of PEs (especially if it is implemented on SIMD machine). Therefore, an O(n) increase in the computation time can be predicted. Table 1 and Figure 7 below illustrate the results of the parallel implementation of DGO on MP-1 and NCUBE.

## Discussion and Conclusion

Parallel implementation of DGO on MP-1 resulted and approximate speed up of 127 times with only 128 allocated processors. This linear speedup with respect to the number allocated PEs is attributed to the insignificant communication time in comparison to the computation time by each PE.

The speedup measured on NCUBE is of O(n) while allocated number of PEs does not exceed 8. Speedup measured with allocation of 16 or more PEs resulted less than O(n) speedup. This phenomenon can be explained by considering the computational power of PEs at each node of NCUBE. Each node of NCUBE contains a powerful 64 bit CPU with a dedicated floating point processor. The existence of this computationally powerful chip significantly reduces the computation time (in comparison to MP-1) while the communication time remains relatively the same. Thus the little communication needed by DGO can no longer be considered to be insignificant (with respect to computation time). Distribution of a job across the less number of PEs will increase the computation time with respect to communication, therefore allowing an O(n) speedup.

While some algorithms reach an execution time of exponential complexity, DGO does no worse than $O(n^2)$. DGO has been found to perform satisfactorily with artificial as well as practical functions. Parallel implementation of this algorithm on MP-1 has demonstrated O(n) speedup. Therefore the parallel implementation of this algorithm exhibits an O(n) in execution time. DGO's success in finding the global optimal point of the problem in combination with its speed of execution sets this algorithm as one the few practical global optimization methods.

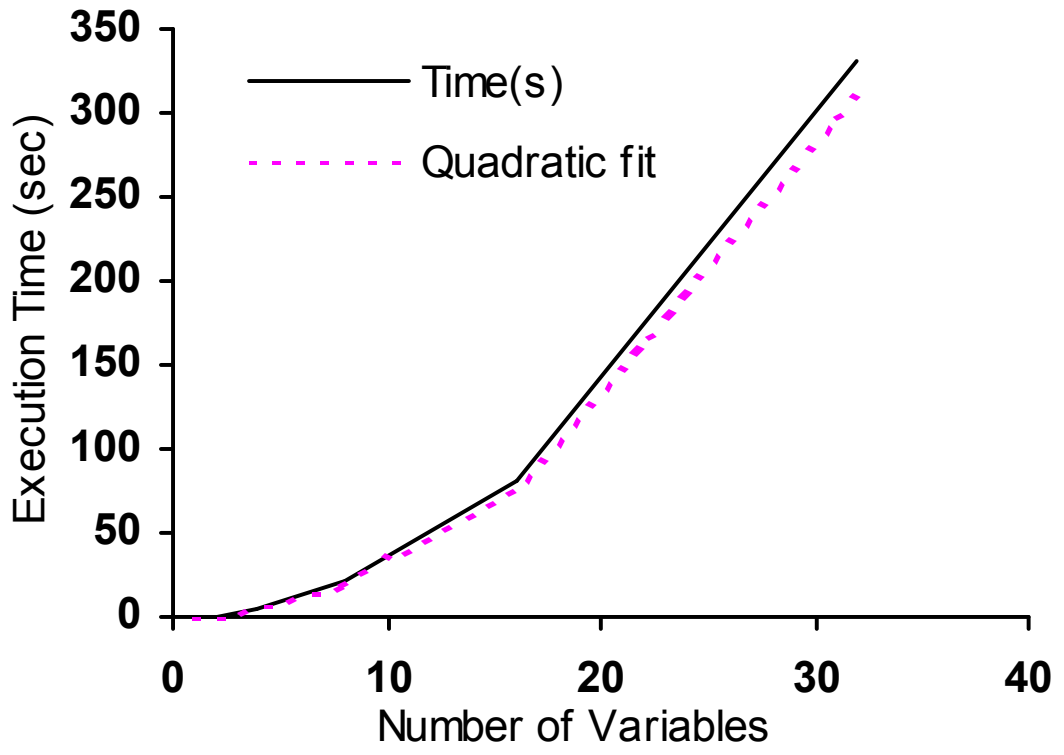

**Figure 6. Sequential execution time for DGO measured on SPARC IV.**

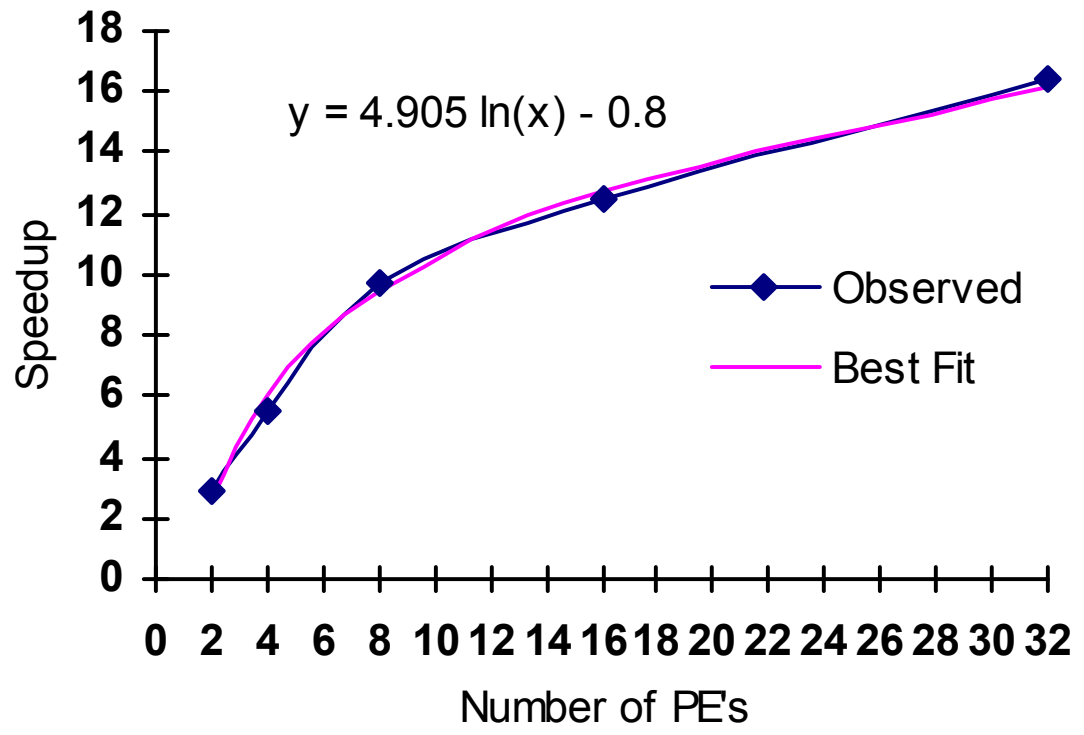

**Figure 7. Speedup (measured in reference to SPARC IV) versus number of PEs on NCUBE.**

**Table 1**. Observed speedup and execution time for different computers and different number of PEs.

| Computer/PEs | Execution time | Speedup with respect to SPARC IV |
|---|---|---|
| SPARC IV | 139.0 | 1.0 |
| MasPar/128 | 1.1 | 126.4 |
| NCUBE/2 | 48.2 | 2.9 |
| NCUBE/4 | 25.4 | 5.5 |
| NCUBE/8 | 14.4 | 9.7 |
| NCUBE/16 | 11.1 | 12.5 |
| NCUBE/32 | 8.5 | 16.4 |